# A new method for microscale cyclic crack growth characterization from notched microcantilevers and application to single crystalline tungsten and a metallic glass

S. Gabel[1,a)] , B. Merle[1,2], E. Bitzek[1,3], M. Göken[1]

[1] Department of Materials Science and Engineering, Friedrich-Alexander-Universität Erlangen-Nürnberg (FAU), Institute I – General Materials Properties and Interdisciplinary Center for Nanostructured Films (IZNF), 91058 Erlangen, Germany
[2] Institute of Materials Engineering, University of Kassel, 34125 Kassel, Germany
[3] Department Computational Materials Design - Microstructure & Mechanics, Max-Planck-Institut Für Eisenforschung, 40237 Düsseldorf, Germany
a) Address all correspondence to this author. e-mail: stefan.s.gabel@fau.de

The lifetime of most metals is limited by cyclic loads, ending in fatigue failure. The progressive growth of cracks ends up in catastrophic failure. An advanced method is presented for the determination of cyclic crack growth on the microscale using a nanoindenter, which allows the characterization of > 10,000 loading cycles. It uses focused ion beam fabricated notched microcantilevers. The method has been validated by cyclic bending metallic glass and tungsten microcantilevers. The experiments reveal a stable crack growth during the lifetime of both samples. The metallic glass shows less plasticity due to the absence of dislocations, but shows shearing caused by the deformation. The crack growth rates determined in the tests follow Paris' power law relationship. The results are reliable, reproducible and comparable with macroscopic setups. Due to the flexibility of the method, it is suitable for the characterization of specific microstructural features, like single phases, grain boundaries or different grain orientations.

## Introduction

The understanding of the fatigue mechanisms is fundamental for the design of mechanical components in most engineering applications. The prediction of the fatigue lifetime is crucial for scheduling the maintenance of structural, transportation and electrical components. The mechanical damage in those cases can be caused by a multitude of loading scenarios. Alternating start-up and shut-down cycles lead to low cycle fatigue (LCF), small vibrations sum up over long time for high cycle fatigue (HCF), thermal cycling leads to thermo mechanical fatigue, etc. Consequently there is a need for different techniques to characterize fatigue under different loadings, environments and scales [1–5]. Microsamples are so far only seldom used to determine the fatigue failure and crack growth behavior. This is a consequence of the experimental challenges for tests at the microscale, as explained below.

The underlying mechanisms for fatigue in crystalline metals are based on the generation, gliding and interaction of dislocations. Even small stresses below the technical yield point can cause dislocation glide. This results in dislocation intersections, source activation and other irreversible mechanisms [5–7]. The summation of those small irreversible deformations leads to fatigue failure. A different kind of fatigue comes into play in metallic glasses, as they do not contain dislocations. Their deformation is based on local densification and the formation of shear transformation zones that can accumulate in shear bands. Several studies showed that the fatigue in those material systems depends more on the initiation of plasticity than on the crack propagation. Hence they have often a smaller lifetime than metals [8–10]. While conventional macroscopic-scale fatigue tests capture the sum of the damaging mechanisms, the influence of specific microstructural features, like the orientation of a grain boundary, on the fatigue lifetime is difficult to determine. Small scale testing allows to focus on specific fatigue crack-microstructure interactions. Consequently those tests can improve the understanding of the fundamental mechanisms contributing to fatigue failure.

Several reliable methods for the characterization of fatigue on the microscale were established and can be used for different geometries and sizes. Thin film methods allow the characterization of < 100-nm-thick samples and enable a direct observation of dislocation structures in situ or ex situ via transmission electron microscopy [11–13]. The downside is that the film thickness requirements limit the application to a small set of material systems. Nanoindenters have a high precision in positioning and applied force, which are suitable for small scale fatigue experiments. However, the normally used pyramidal tip geometry is poorly suited for those experiments, due to the complex stress state and the growing activation zone with indentation depth [14]. Simpler geometries can be tested with custom tips. Fully reversible low cycle fatigue experiments on microbeams were realized with a claw gripper [7, 15–17]. Another approach is to attach the microcantilever to a tungsten tip with electron beam curing glue [18]. These approaches require in-situ setups, precise positioning and have been limited to LCF experiments. A new approach is to utilize the continuous stiffness measurement (CSM) method, that is usually used in nanoindentation systems for monitoring the variation of the hardness and Young's modulus [19]. The high frequencies allow to reach the HCF regime during experiments. This was either used for fully reversible in-situ experiments, as shown by Lavenstein et al. [18], or for more easily accessible ex-situ tests [20, 21]. The advantage of those ex-situ approaches is the less costly setup, but the tests are limited to pure cyclic compression of micro-pillars or cyclic bending of microcantilevers.

In these studies, the characterization on the microscale evidenced fatigue mechanisms and lifetime predictions similar to macroscopic testing [18, 20, 21]. Independent of the initial microstructure (ranging from ultra fine grained to single crystalline metals), those test showed that the failure of the samples is dominated by the formation of extrusions at the surface. Those extrusions grow until they initiate cracks, which result in the final fracture of the microscopic samples. [16–18, 20, 21].

Due to the dimensions of the extrusions and the samples, large fractions of the samples are affected by them. Consequently, the transformation zone to regular crack growth during fatigue is limited. Especially the initial cracks form and grow along the slip planes of the extrusions. As the slip planes are inclined to the loading direction, the fracture mode of the cracks are a mixture of crack mode I, II and III. A full transformation to mode I, as seen on macroscopic fatigue experiments cannot be reached, as the samples fail earlier [16, 18, 21]. Additionally the early cracks have a complex shape, which prevents a good description in terms of the fracture mechanics during microscopic fatigue experiments, similar to macroscopic experiments [6]. However, stable cyclic crack growth is a fundamental part of macroscopic fatigue and is described by Paris' law, shown in Fig. 1(a) [22]. Up to now it has not been observed in micromechanical fatigue experiments. The aim of this work is to demonstrate that Paris' law holds true on the microscale and to investigate its size dependence. To achieve this, a new approach to study cyclic crack growth by cyclic loading of notched microcantilevers [Fig. 1(b)] is used.

## Results and discussion

### Monotonic testing

The quasi-static deformation behavior of the notched microcantilevers was first investigated, to determine the fracture toughness of the materials. Figure 2(a) shows the force versus the displacement signal during the loading of the microcantilevers. Both materials show a linear-elastic initial behavior. The maximum force is nearly twice as high with metallic glass as with W, with nearly the same cantilever dimensions. The later stage of the tungsten cantilever is dominated by apparent strain softening. In contrast the metallic glass sample shows limited strain softening, accompanied by serrations. The metallic glass test was stopped at a displacement of 4.5 μm. The measurements reveal the different behavior of the materials at microscale. While tungsten shows a clear cleavage fracture during the experiment, the metallic glass shows more plasticity. The continuous crack growth resistance was calculated for both materials, as described in literature [23, 24] in Fig. 2(b). Tungsten shows a brittle material behavior, where the crack grows through the whole microcantilever. The initial linear crack blunting stage is followed by faster crack growth. This is similar to results for {001}<100> oriented tungsten microcantilevers [25, 26]. As can be seen for the metallic glass sample in Fig. 2(b) the crack growth resistance stagnates at a crack growth length of 300–350 nm. This explains the stagnation of the force level in Fig. 2(a). Hence the microcantilever showed larger deformations, which prevented further crack growth. Additionally, the abnormal crack growth resistance shows that the fracture toughness of the metallic glass is too high for microcantilever tests. Consequently the calculated fracture toughness is invalid. The different behavior can also be seen in the fracture surfaces in Fig. 3(a, b). The metallic glass cantilever has only a small fractured area, which is ca. 350 nm long, in good agreement with the calculated values. The area below the crack tip shows several shear bands, as common for metallic glasses [9], those shear events correspond to individual small peaks in Fig. 2(a). The shearing at the lower end of the cantilever leads to buckling, which reduces the effective stress intensity. The buckling prevents further crack growth, leading to the higher crack resistance of the metallic glass. The fracture surface of the tungsten cantilever shows that the crack grew through nearly the whole cantilever. The region around the initial notch shows some necking and gliding events. Otherwise there are no surface signs of shearing or buckling. The metallic glasses microcantilevers fracture surface seem to have a slightly higher roughness than the tungsten microcantilever ones. The fracture toughness of the tungsten microcantilever, calculated by the criteria of Ast et al. [24, 26], is 7.2 MPa$\sqrt{m}$ and the one of the metallic glass is 8.9 MPa$\sqrt{m}$. As mentioned above the abnormal shape of the crack growth resistance curve already shows that the value for the metallic glass is invalid due to its plastic deformation. Hence, the fracture toughness of the metallic glass is smaller than in macroscopic studies (16.9 ± 1.5 MPa$\sqrt{m}$) [10]. The results of tungsten fit well to {001}<100> oriented macroscopic studies for with a fracture toughness of 6.2 ± 1.7 MPa$\sqrt{m}$ [27].

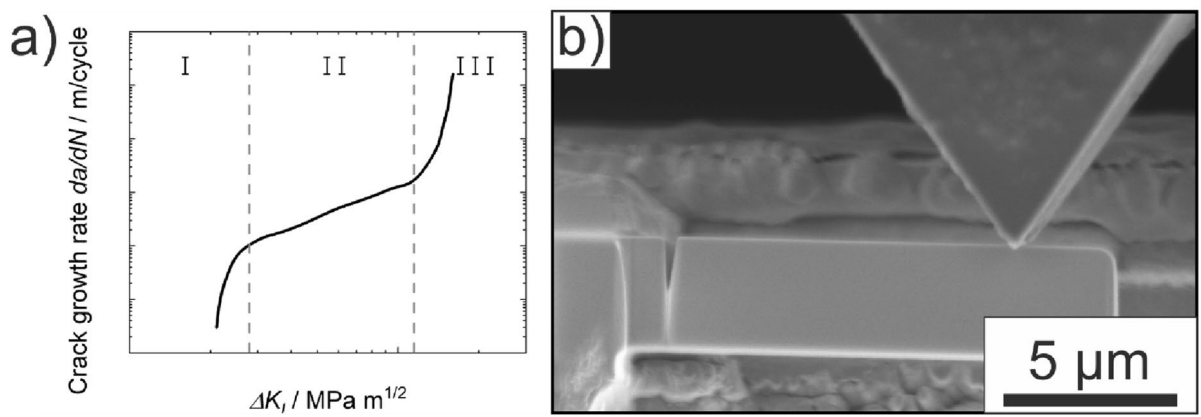

**Figure 1:** (a) Sketch of Paris' plot. (b) SEM micrograph of a metallic glass microcantilever during microbending.

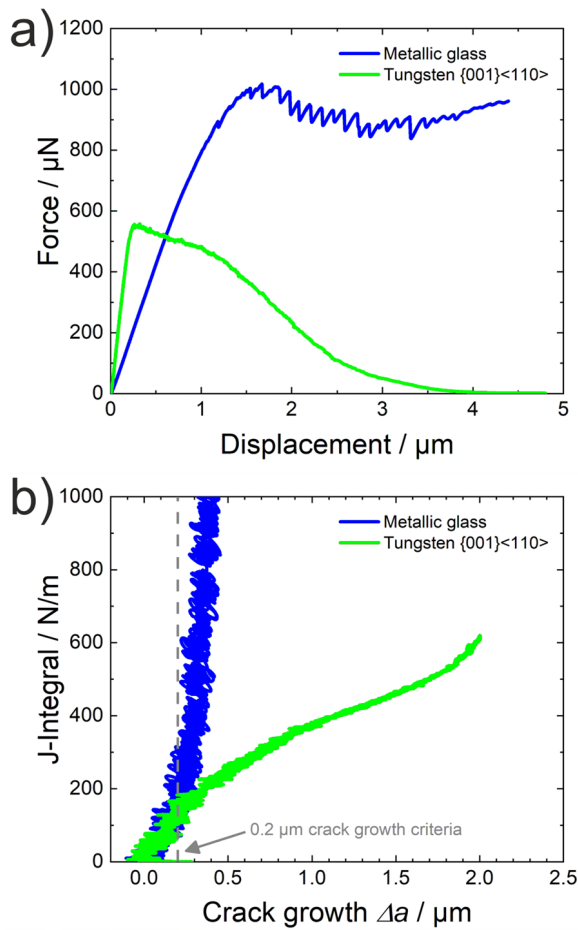

**Figure 2:** Monotonic bending results for the metallic glass and tungsten single crystal microcantilevers. The tungsten samples had a {001}<110> crack orientation. (a) Force–displacement curve. (b) Continuous crack resistance curves.

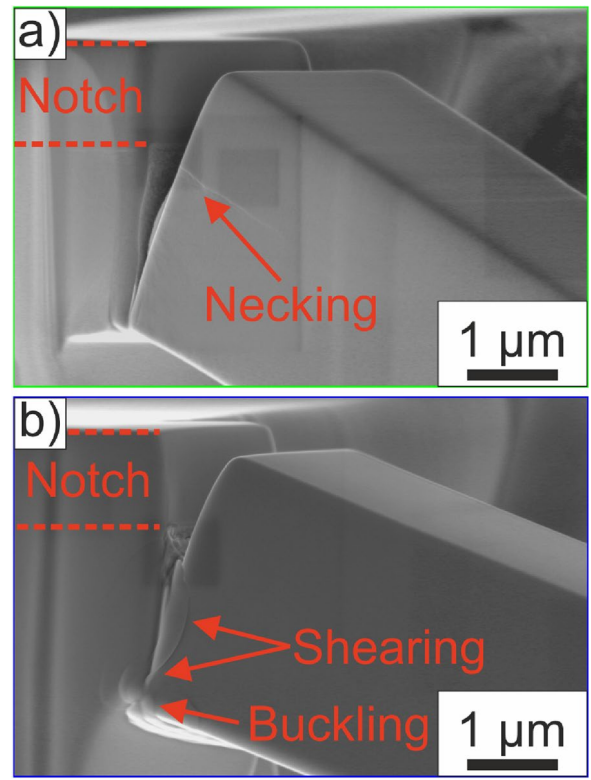

**Figure 3:** Electron micrographs of the (a) tungsten and (b) metallic glass monotonic fracture surfaces post testing. The metallic glass cantilever shows deformation by shear bands and buckling at the lower end of the cantilever, which stopped further crack growth. The tungsten sample shows necking and the crack propagated through the whole microcantilever. The cantilevers are tilted by 45°.

### Evolution of the experimental parameters during cyclic testing

Figure 4 shows the cyclic parameters for a metallic glass microcantilever that was recorded at 400 Hz and an R value of 0.1. In Fig. 4(a) the force–displacement cycles are shown for different stages of the experiment, ranging from 10 to 95% of the lifetime. The lowest displacement point in each cycle was set to be zero, to compensate for the progressive overall displacement of the cantilever during the test. Different trends are directly observable. First the maximum force in each cycle decreases, while the maximum relative displacement increases with higher cycle numbers. As the experiments were set to have a constant force range of $\Delta P$ = 446 µN controlled by a feedback loop of the natively displacement controlled indenter, the controller of the indenter is too slow at regulating the force levels at the end of the tests. The force range is shown in Fig. 4(b) and it also indicates this decrease at the end of the lifetime.

Second, in Fig. 4(a) the slope of the load–displacement cycles decreases during the measurements. That indicates that the stiffness of the cantilevers is reduced, which is a result of crack growth. The CSM stiffness decrease, measured by the indenter, is also shown in Fig. 4(c). In comparison, we calculated the unloading stiffness $S_{unload}$ from the unloading part of each cycle, by fitting a linear regression to the area between 100 and 70% of the maximum load. This represents the elastic material response. The unloading stiffness should be higher than the CSM stiffness, as it does not contain the plastic deformation. However it is lower due to the bended shape of the loading cycles in Fig. 4(a). Ideally, those cycles should have an elliptical shape. We assume that the shape is an overlay effect from the indentation of the indenter tip into the cantilever. A correction of this effect was not possible, because the contact area between tip and cantilever also changes with ongoing cycling. This results in an increasing contact depth, which would not be described by single indentation experiments. A normal cyclic indentation study in H-bar shaped structures with the same dimensions would neglect the effect that the microcantilever bends during the experiment. Hence non orthogonal cyclic indentations would be required. For a full reconstruction of the mentioned effects, each microcantilever would require several further measurements, reducing the effectiveness of the method. Nevertheless, the stiffnesses mainly differ in their value, but their evolution during the experiments is nearly similar, as shown in Fig. 4(c). The relative difference is nearly constant with ~ 4%, as shown in Supplementary Information Figure S1. This difference is irrelevant for the calculation of the crack length, as the conversion calculated by FEM [28] only uses the relative stiffness of the cantilever. Hence it can be concluded, that the CSM stiffness is an acceptable and easier method for the calculation of the crack growth.

The change of the crack length with the number of loading cycles is given in Fig. 4(d). It was calculated from the CSM stiffness and it is continuously increasing. The crack growth per cycle is also increasing, which causes an accelerative failure at the end of the experiments, as shown in Fig. 4(e). The curve is a locally weighted polynomial regression smooth [29], which was used, to reduce the noise of the CSM signal. There is still some noise visible within the oscillations between 400 and 800 cycles. The maximum crack growth was ~ 800 nm, when the experiment was stopped, due to the wedge indenter losing contact to the mircocantilever. Considering the initial notch depth of ~ 1 µm and the cantilever height of ~ 3 µm, the crack grew through a third of the cantilever thickness before failure.

Using the actual crack length and the force range in each cycle, the stress intensity factor range in each cycle can be calculated by formula 2. The results are shown in Fig. 4(f)). The stress intensity factor increases permanently with an increasing rate due to the corresponding crack growth. This compensates the instrument-caused force range reduction in Fig. 4(b) at the end of the test. As the stress intensity and the crack length are both increasing during the experiment, the requirements for the Paris law are fulfilled [30]. A similar analysis of the cyclic parameters for a tungsten microcantilever is shown in Supplementary Information Figure S2. Except for the full force–displacement cycles, which were not recorded with the highest data acquisition rate, the cyclic response is similar. An obvious difference between both materials is that the cyclic deflection of the tungsten cantilevers is smaller at similar forces, due to their higher elastic modulus.

### Morphology of the cyclic tested fracture surfaces

After testing, the fatigued specimens were imaged to observe the fracture surface. Postmortem images corresponding to the different samples are shown in Fig. 5. The metallic glass microcantilever, which is described in detail in Fig. 4, reveals different stages of the cyclic crack growth in Fig. 5(a). The initial fracture area under the initial notch shows a rough surface with several lines, which represent striations. Additionally an increase of the striation size with higher cycle numbers is observable. The first ones under the initial notch are below the microscope resolution, whereas the last ones have a size > 10 nm. The length of those striations is in good accordance to the determined crack growth rate of Fig. 4(e), which are also in the range of ∼ 10 nm. This area with striations has a length of ∼ 300 nm. At a longer crack length the fracture surface has a higher roughness and the line structure disappears. The crack is stopped after a length of ~800 nm, which is in good agreement with the calculated crack growth in Fig. 4(d). The microcantilever is still connected to the root and the remaining thickness shows shearing. This shearing caused the plastic deformation of the cantilever, which resulted in the indenter tip losing contact. The experiment was stopped accordingly.

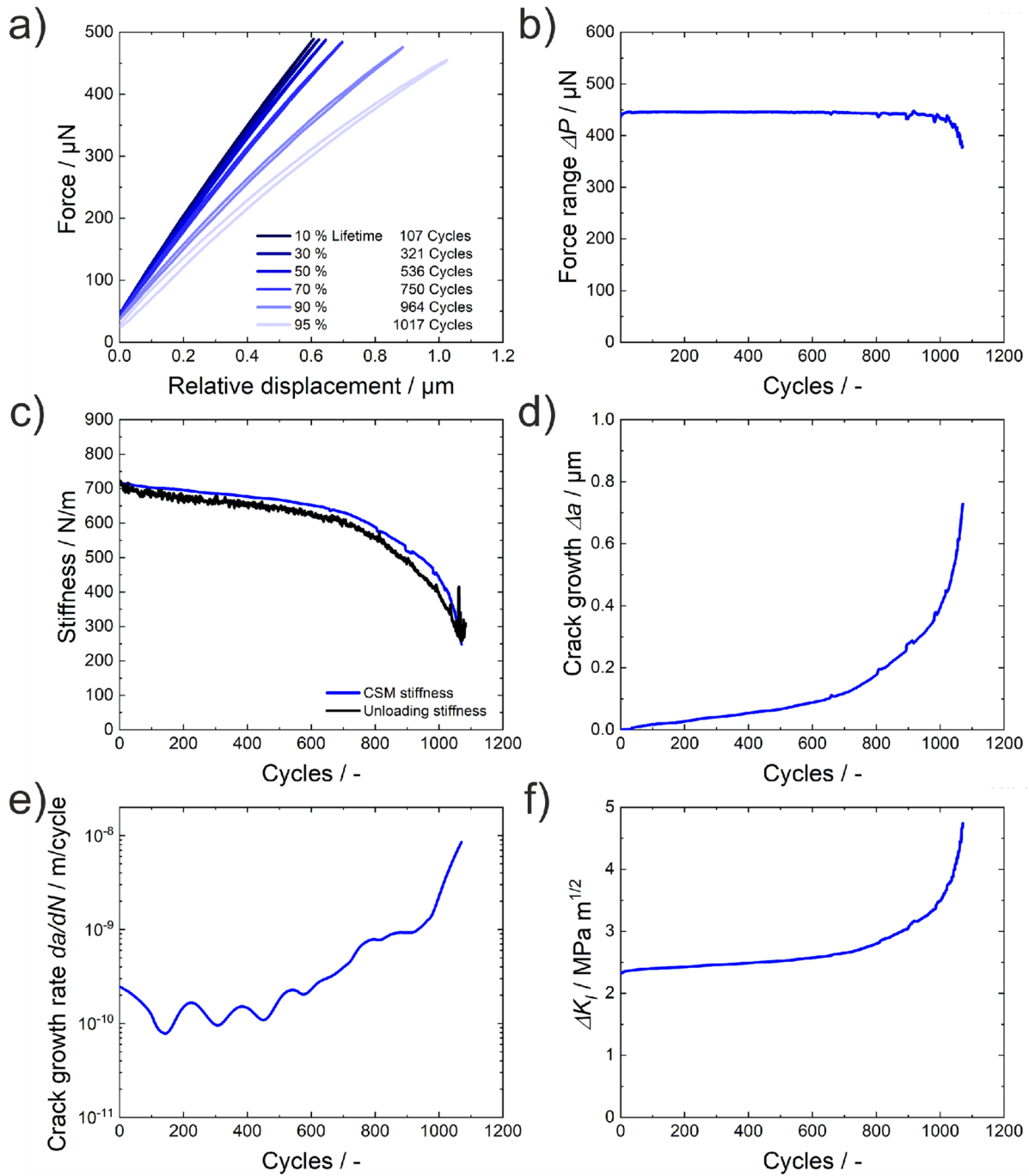

**Figure 4:** Cyclic results of a metallic glass microcantilever. (a) Cyclic force–displacement signal at different stages of the fatigue experiment. (b) Nearly constant force range throughout the fatigue tests for $R = 0.1$. (c) Comparison between the stiffness measured by the indenter and the stiffness calculated from the unloading cycles. (d) Evolution of the crack length during the fatigue cycles. (e) Crack growth rate during the experiments. (f) Evolution of the stress intensity factor range.

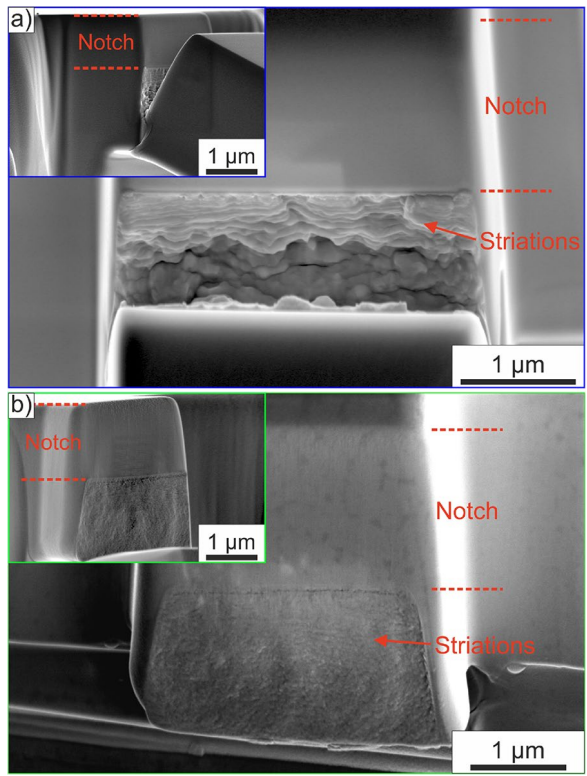

**Figure 5:** SEM images of the fracture surfaces of different microcantilevers after cyclic fatigue crack growth experiments. (a) Metallic glass cantilever with the higher recording frequency, described in Fig. 4. (b) Tungsten microcantilever with a {001}<110> crack orientation. The SEM images show the 45° tilted fracture surface. The insets show a view direction inclined to the side of the cantilever.

The tilted fracture surface of a tungsten microcantilever in Fig. 5(b) shows a different behavior than the metallic glass. Below the initial notch, striations are also observable, but they have a smaller spacing than in the metallic glass. Supplementary Information Figure S3 shows fracture surface images with higher magnification, in which the striations are better visible. The dimples, which formed during the crack growth are smaller than in the metallic glass. Hence, the transformation to the more plastic failure at the lower end of the fracture surface is more gradual. The crack grew through the entire microcantilever and only a small rest connects the beam to its root.

The tests show no signs of early stage fatigue mechanism, like extrusion formation and micro cracking [7, 21, 31]. The cracks grow by small steps during each cycle until they are arrested. The size of the so formed striations increases. Consequently the crack surfaces in both materials share the same fatigue features as seen in macroscopic tests [6, 32].

**Evolution of cyclic parameters**

The cyclic parameters presented in Sect. 2.2 were used to measure the cyclic crack growth for each tested microcantilever. The results are given in Fig. 6. The crack growth is shown in Fig. 6(a, b) in dependence on the cycle number for the metallic glass and the tungsten samples. The bright colors encode tests with a higher initial stress intensity range. All curves show the same trend: the crack length as well as the crack growth rate increase with higher cycle numbers. Some tests end after a rather small crack growth of ~0.75 μm, which is equal to ca. 30% of the cantilevers unnotched thickness $W$-$a$. In this case, the indenter tip lost contact during the experiment, which was mainly seen in the metallic glass samples. Furthermore, higher initial stress ranges shorten the lifetime of the samples and the stress intensity range increases with higher cycle numbers and the slope of the graphs rises in Fig. 6(c, d). This trend is expected from Paris' law [22], which states that an increase of the crack length increases the stress intensity and vice versa, for constant load amplitudes. As explained in Sect. 2.2, the small force range drop at the end of the experiment is compensated by the crack growth. To correlate the crack growth with the stress intensity range both parameters are shown on a logarithmic scale in Fig. 6(e, f). The curve is a locally weighted regression smooth [29], which was used, to reduce the noise of the crack length calculated from the CSM stiffness. For both materials the plots show the transition from the very low initial cyclic crack growth in regime I to the nearly linear relation in regime II. In the regime I some tests show an initially higher crack growth rate, which then decreases to rise again later. It can be assumed that this the influence of the focused ion beam (FIB) preparation and the formation of a plasticity-induced crack closure zone. The initial crack area is highly damaged and implanted by Ga Ions, which reduces the fracture toughness locally [33]. The size of this damage zone depends on the material and is smaller than 70 nm, as shown by several studies [24, 33, 34]. After the crack grows past this region, it stabilizes and shows the fatigue behavior of the pristine material.

The initial crack growth rate per cycle in regime I is between $10^{-11}$ and $10^{-10}$ m/cycle in both materials, as shown in Fig. 6(e–f). This is in the lower range of macroscopic studies and allows to define the threshold toughness $\Delta K_{TH}$, which was defined as the $\Delta K_I$ for which the growth rate is $< 10^{-10}$ m/cycle in ASTM standard E647 [30, 35–37]. $\Delta K_{TH}$ is $2.5 - 2.8$ MPa$\sqrt{m}$ for the metallic glass and is $2.2 - 2.5$ MPa$\sqrt{m}$ for the tungsten single crystal. This threshold was also observed in a pre-test, where a tungsten single crystal microcantilever was loaded with an initial stress intensity range of 2.0 MPa$\sqrt{m}$. The cantilever withstood 100 000 cycles without visible crack growth and only a small drop of the stiffness (Supplementary Information Figure S4. The transition to regime II in macroscopic studies [35–37] is also between $10^{-9}$ and $10^{-8}$ m/cycle, which is in accordance to the microscale experiments.

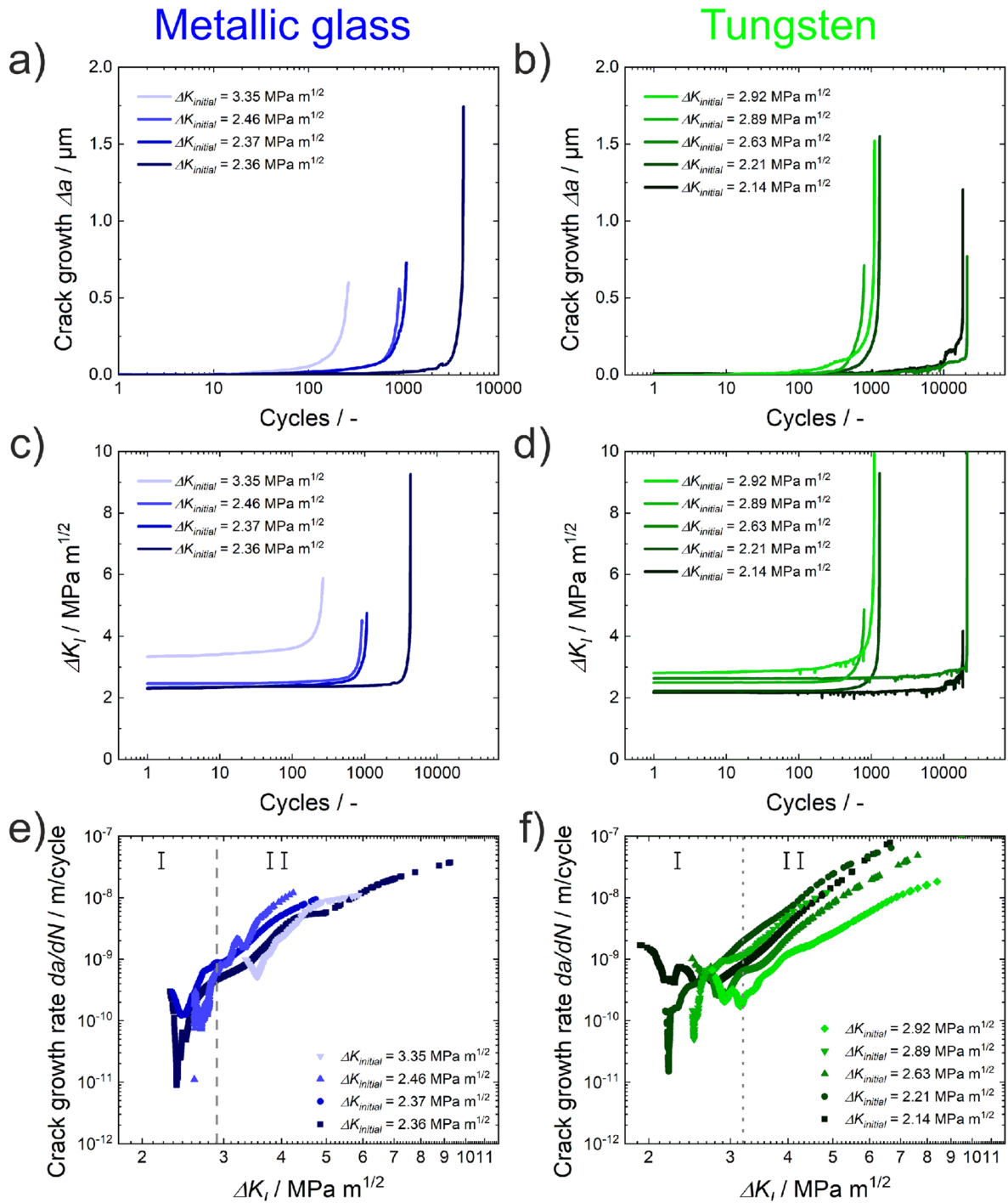

**Figure 6:** Cyclic parameters of the microcantilever fatigue experiments of the metallic glass (blue) and tungsten (green). (a, b) The increase of crack length with higher cycle numbers. (c, d) Evolution of the stress intensity range. (e, f) Plot of the crack growth rate per cycle with respect to the stress intensity range.

After the crack grows through regime *I*, it stabilizes. Additionally there is a transition in the type of crack. In the initial state, the crack can be described as a mechanically short crack [38], as there are no pronounced crack closure effects due to the unfatigued notch. With the start of cyclic testing, a plasticity-induced crack closure can form. This reduces the crack growth rate and initiates a *K*-determined crack-propagation behavior as a physically short crack [38]. This regime II is described by the Paris' power law relationship [22]:

$$\frac{da}{dN} = C\Delta K_I^m, \quad (1)$$

where *da/dN* is the crack growth rate, $\Delta K_I$ is the stress intensity, while *C* and *m* are the scaling constants. The Paris' equation exponents were found to be $m = 3.8 \pm 1.7$ and $C = 1.5 \times 10^{-11} \pm 0.4 \times 10^{-11}$ for the metallic glass. While the ones for the tungsten single crystals are $m = 4.3 \pm 0.8$ and $C = 6.7 \times 10^{-12} \pm 3.5 \times 10^{-12}$.

The Paris' equation exponents can also be compared to macroscopic results. However, there are no macroscopic fatigue tests for the metallic glass $Zr_{48}Cu_{36}Ag_8Al_8$. Hence it is compared to other ZrCu based metallic glasses, which have comparable material properties. These studies determined a stress exponent *m* of 1.7–3 and a $\Delta K_{TH}$ of $1.8 - 2$ MPa$\sqrt{m}$ [8, 35, 39], which is slightly lower than the $m = 3.8$ and $\Delta K_{TH}$ is $2.5 - 2.8$ MPa$\sqrt{m}$ for the microscale tests. Studies on cold rolled tungsten measured an *m* of 2.8, which is also below the 4.3 of our results [36]. The $\Delta K_{TH}$ in this study with ∼ 10 MPa$\sqrt{m}$ also five times higher than in the microscale tests. The reasons for these small differences are the different materials composition and microstructure. The metallic glasses of the other studies were shown to have a higher plasticity than $Zr_{48}Cu_{36}Ag_8Al_8$, which also increases the fatigue life. Cold-rolled tungsten leads to a ductilization [27, 36], which increases the crack growth resistance in comparison to single crystals. Additionally, the investigated {001} <110> orientation is a weak one, which has a lower fracture toughness than other directions [27].

### Reliability of the dynamic microcantilever method

The evidenced cyclic crack growth behavior in the Paris' plot is similar to macroscopic literature data [6, 22, 32]. The cracks grow stable with an increasing speed. The crack growth rate transitions to a stable region is described by the Paris' law. The limitation to positive R ratios with tensile stresses in microcantilevers prevents full crack closure and could bring along cyclic creep effects [40]. The creep by the mean stresses should increase the crack growth speed. However, creep effects can be neglected for the tested specimens, because of the low homologous temperature during the experiments and the small dislocation activity of tungsten as well as no dislocations in the metallic glass at those temperatures [25].

Crack growth was observed through the whole microcantilevers. According to Fig. 6 the cyclic crack growth consisted only of regime I and II. Those regimes only describe the onset of crack growth and stable crack growth [22]. The regime III failure by rupture is missing. This is a consequence of the sample dimensions and the bending geometry, which is already known from previous microcantilevers fracture studies. Clamping effects at the root of the cantilever stabilize the crack, before reaching the other end of the cantilever [24, 41, 42]. The forces preventing further crack growth at the bottom surface of the microcantilever have different origins. Due to the bending geometry, the microcantilever contains a stress gradient with compressive stresses at the bottom. This prevents the development of a plastic zone throughout the microcantilever, e.g. through unconstrained dislocation motion or shear band formation. Only when the remaining ligament of the cantilever becomes very small, does the plastic zone below the crack tip expand through the whole residual thickness. This corresponds to a change to plane stress, which also inhibits further crack growth [26]. The formation of a long crack, as described for fatigue, cannot happen, as it would have to be > 0.5 mm [38]. This also does not allow the change to regime III. Further the contact of the tip to a constant loading position on the microcantilever is only guaranteed by friction. Hence the tip could slide off the contact point during testing. Measurements of the span length between the loading point and the notch in the in situ videos show no horizontal movement of the tip (see Online Resource 1 and Online Resource 2). Additionally, post mortem images also show no signs of tip sliding, as shown in Supplementary Information Figure S5. The contact loss in the final stage of testing is an effect of the indenter feedback loop. The feedback of the displacement-controlled indenter is too sluggish to cope with the fast increase in cyclic displacement towards the end of the experiments. As a result, the minimal force during a cycle $F_{min}$ drops below its setpoint so that the tip loses contact. A workaround would be to use intrinsically force controlled indenters. Furthermore the influence of the tip's contact area on the cyclic load–displacement behavior needs to be further investigated, as described in chapter 2.2. It is expected to highly depend both on the tip geometry and the tested material. This would give a better control over the stiffness and hence the crack length. The usage of clamped actuators and pure tensile tests, as used by other groups [7, 15–17], could overcome those effects. On the downside, this would increase the experimental complexity.

The Paris' law [22] relations in Fig. 6(e, f) are consistent for the metallic glass microcantilevers. Regardless of the initial stress intensity factor the different experiments tend to overlap in the linear section of regime II. The same can be seen for the tungsten microcantilevers but with a higher deviation in the results. Nevertheless, the stress exponents are in the same range as in macroscopic results, which shows the reliability of the method and its comparability to macroscopic tests.

An advantage of the method is that it could be used ex-situ, which allows faster and cheaper testing, as ex-situ nanoindenters are more common. They also allow the characterization of atmospheric influences. The good control over the crack length by fatigue could be used to create fatigued notches for fracture toughness measurements with microcantilevers. This would overcome the issue of blunt notches obtained from FIB milling, which have a radius of ~ 10 nm [33, 42]. Additionally those fatigued notches would comply with the requirements of ASTM norms [30, 43].

## Conclusions

The results show that the high cycle fatigue testing method of FIB-milled microcantilevers by CSM can be adjusted to measure cyclic crack growth. The method allows to determine the evolution of the crack propagation and the stress intensity range during the experiment. This enables to evaluate the transition from regime I to regime II crack growth, which allows to track the stable crack growth regime described by Paris' law.

The notched microcantilever geometry was used to investigate cyclic crack growth in a metallic glass Zr48Cu36Ag8Al8 and tungsten single crystal in the {001}<110> crack orientation. During testing both materials showed the formation of striations on their fracture surfaces. The resting line distance increased with higher cycle numbers and higher stress intensity ranges. The crack growth resistance in monotonic experiments was higher for the metallic glass, due to shear banding as an effective mechanism for crack stopping. Accordingly, the fatigue tests showed a higher cyclic crack growth resistance of the metallic glass samples. The stress exponents in the Paris' law were comparable to literature results of macroscopic tests.

Following the present validation of the method, it could be used in the future to characterize the cyclic crack growth resistance of microstructural constituents, such as individual phases and grain boundaries.

## Material and methods

### Material and sample preparation

For the experiments two materials were chosen. One of those materials was the metallic glass $Zr_{48}Cu_{36}Ag_8Al_8$. The samples were made by centrifugal vacuum casting, in which the 1300 °C hot mold was quenched in a copper mold, as described by Haag et al. [44]. The metallic glass is perfectly isotropic and has no crystallograhpic cleavage planes that could redirect the cracks. As the interaction between cracks and dislocations is fundamental for the fatigue of metals, also a pure tungsten single crystal (purity 99.999%) was investigated. The single crystal was obtained from *MaTecK Material-Technologie & Kristalle GmbH, Germany*. The samples were subsequently grounded and electropolished to get a deformation free surface area.

Pre-notched microcantilevers were milled on the edges of the samples by FIB milling in a Helios Nanolab 600i (*ThermoFisher Scientific, MA, USA*). The first ion beam cuts were made with an acceleration voltage of 30 kV and a current of 9.3 nA. In further steps, the current was progressively reduced to 780 pA. A small notch radius of 10 nm was accomplished by milling the notch with a current of 80 pA, as shown in previous studies [42, 45]. The final dimensions of the cantilevers were measured via scanning electron microscopy (SEM) and had an aspect ratio of 10:3:3 μm (*L*(*length*):*W*(*width*):*T*(*thickness*)) with a crack notch length $a_0$ of ~ 1 μm as shown in Fig. 1(b). In total four metallic glass and five tungsten microcantilevers were made for testing. As the tungsten samples were made on a single crystal the orientation was determined by Electron back scatter diffraction (EBSD). The crack orientation was {001}<110>. Per convention, the first parameter describes the crack plane and the latter represents the crack front direction.

### In situ testing

The microcantilevers were loaded with a displacement-controlled in situ indenter (*NMT 03, FemtoTools, Switzerland*) inside a SEM (*1540EsB, Carl Zeiss, Germany*) at room temperature. The indenter was equipped with a 10-μm-long wedge tip. The cyclic method of Gabel and Merle [21] was adapted for the testing of notched microcantilevers. The experiments are load controlled and the stress intensity factor $K_I$ was calculated from the constant force range [46]:

$$K_I = \frac{6PL\sqrt{\pi a}}{TW^2} f\left(\frac{a}{W}\right), \qquad (2)$$

where $P$ is the loading force, $L$ is the span length between the loading point and the notch, $T$ is the beam thickness, $W$ is the beam width and $a$ the crack length. $f(a/W)$ is a dimensionless geometry function of the tested specimens, which was described by Riedl et al. [46]. The maximum stress intensity in each cycle of the experiments is $K_{I,max}$. To avoid the loss of contact between the tip and the cantilevers during the oscillations, the unloading segment ended with a stress intensity factor $K_{I,min}$, which was 10% of $K_{I,max}$. Hence the stress intensity ratio $R = K_{I,min}/K_{I,max}$ was 0.1 at the beginning of all experiments.

The test were started by loading the cantilevers to the mean stress intensity $(K_{I,min} + K_{I,max})/2$. Then the 10 Hz CSM oscillation was ramped-up within 25 cycles to the initial $\Delta K_I$. The first cycle was defined when $\Delta K_I$ was reached. To track the cantilevers during the experiments, movies were recorded by SEM. To overcome the vibrations from actuation of the cantilevers during the SEM recording, the imaging time was set to a multiple of 100 ms. The difference by non-matching imaging frequencies can be seen in Online Resource 1 and Online Resource 2. Both Online Resources show fatigue tests of tungsten microcantilevers. The first test has an image recording time of 111 ms, while Online Resource 2 has a recording time of 1000 ms. The second one has a better image quality and the microcantilever image stays nearly constant over a long time period. Tests ended either when the indenter tip lost contact, due to a too large deflection of the microcantilevers, which could be also a result of the feedback loop being too sluggish to stop the cantilever from losing contact, or when the microcantilevers broke off. For each cycle one data point per measured and saved, which was generated by the higher recording rate of the instrument. After the tests the fracture surfaces were inspected via SEM.

The experiments used a feedback loop of the indenter for maintaining a constant load range $\Delta P$. This leads to an increase of the cyclic stress intensity, if the initial notch grows. The crack growth $\Delta a$ can be determined by the change in dynamic cantilever stiffness, from the CSM signal. The CSM stiffness $S_{CSM}$ is calculated as the ratio between the maximal and minimal force and displacement in each loop. The stiffness and $\Delta a$ are correlated, as shown via Finite Element (FE) modeling [28]. The prerequisite for this model is that the plastic deformation in each cycle is negligible. For simplification this was assumed for all cycles. As the crack grows, the stress intensity as well as the plastic deformation in each cycle increase. Hence the stiffness will get underestimated during the experiments. In one test the acquisition rate of the indenter was set to 400 Hz to get 40 data points for each fatigue cycle $N$ and to determine the effects in each cycle. The mean force for this experiment was set to be 273 µN and the force range to 446 µN. With those parameters the initial stress intensity range, as calculated with Eq. 1, was $\sim 2.37$ MPa$\sqrt{\mathrm{m}}$ with an $R$ value of 0.1.

For comparison between the cyclic behavior and the monotonic behavior of the materials, additional monotonic loading tests of notched microcantilevers were done. One microsample was measured per material in the in situ indenter. The displacement rate of the tip was set to 20 nm/s. The evaluation methodology for those fracture toughness measurements are described elsewhere [23–25]. To determine the fracture toughness of the materials, the 0.2 µm crack growth criteria was used, which was established by Ast et al. [24, 26]. It describes the fracture toughness as the intersection of the crack growth resistance with a vertical line at 0.2 µm crack growth. The criterion only describes the fracture initiation, which causes an underestimation for materials with high plasticity at larger crack growth. However, it can be used for brittle and semi-brittle materials and its results are in line with macroscopic experiments.

### Acknowledgments

This project has received funding from the European Research Council (ERC) under the European Union's Horizon 2020 research and innovation programme (microKIc, Grant agreement No. 725483 and NanoHighSpeed, Grant agreement No. 949626). This research used resources from the Center for Nanoanalysis and Electron Microscopy (CENEM) and the Interdisciplinary Center for Nanostructured Films (IZNF) at Friedrich-Alexander University Erlangen-Nürnberg (FAU).

### Author contribution

SG: conceptualization, methodology, investigation, writing—original draft, BM: resources, formal analysis, writing—reviewing. EB: writing—reviewing. MG: writing—reviewing.

### Funding

Open Access funding enabled and organized by Projekt DEAL.

### Data availability

The datasets generated during and analyzed during the current study are available from the corresponding author on reasonable request.

### Declarations

**Conflict of interest** On behalf of all authors, the corresponding author states that there is no conflict of interest.

### Supplementary Information

The online version contains supplementary material available at https://doi.org/10.1557/s43578-022-00618-x.